\documentclass[12pt]{iopart}

\expandafter\let\csname equation*\endcsname\relax
\expandafter\let\csname endequation*\endcsname\relax

\usepackage{graphicx,color}
\usepackage{amsmath,amssymb,latexsym}
\usepackage{mathtools}
\usepackage[english]{babel}
\usepackage{dcolumn}
\usepackage{bm}
\usepackage{float}
\usepackage[normalem]{ulem}

\usepackage{color}

\begin{document}

\title[Brownian non-Gaussian polymer diffusion]{Brownian non-Gaussian polymer diffusion and queing theory in the mean-field limit}

\author{Sankaran Nampoothiri, Enzo Orlandini, Flavio Seno, Fulvio Baldovin}
\address{Dipartimento di Fisica e Astronomia `G. Galilei' - DFA, Sezione INFN,
Universit\`a di Padova,
Via Marzolo 8, 35131 Padova (PD), Italy} 
\ead{sankaran.nampoothiri@unipd.it}
\ead{fulvio.baldovin@unipd.it}
\ead{orlandini@pd.infn.it}
\ead{flavio.seno@unipd.it}

%\author{Sankaran Nampoothiri}
%\email{sankaran.nampoothiri@unipd.it}
%\affiliation{ Dipartimento di Fisica e Astronomia `G. Galilei' - DFA, Sezione INFN, Universit\`a di Padova, Via Marzolo 8, 35131 Padova (PD), Italy}

%\author{Enzo Orlandini}
%\email{orlandini@pd.infn.it}
%\affiliation{ Dipartimento di Fisica e Astronomia `G. Galilei' - DFA, Sezione INFN, Universit\`a di Padova, Via Marzolo 8, 35131 Padova (PD), Italy }

%\author{Flavio Seno}
%\email{seno@pd.infn.it}
%\affiliation{ Dipartimento di Fisica e Astronomia `G. Galilei' - DFA, Sezione INFN, Universit\`a di Padova, Via Marzolo 8, 35131 Padova (PD), Italy }

%\author{Fulvio Baldovin}
%\email{baldovin@pd.infn.it}
%\affiliation{ Dipartimento di Fisica e Astronomia `G. Galilei' - DFA, Sezione INFN, Universit\`a di Padova, Via Marzolo 8, 35131 Padova (PD), Italy }

\date{\today}

\begin{abstract}
  We link the Brownian non-Gaussian diffusion of a polymer center of mass
  to a microscopic cause: the
  polymerization/depolymerization phenomenon occurring when the polymer
  is in contact with a monomer chemostat.
  The anomalous behavior is triggered by the polymer critical point,
  separating the dilute and the dense phase in the grand canonical
  ensemble.
  In the mean-field limit we establish contact with queuing theory and
  show that the kurtosis of the polymer center of mass diverges alike a
  response function when the system becomes critical,  
  a result which holds for general polymer dynamics (Zimm, Rouse,
  reptation).
  Both the equilibrium and nonequilibrium behaviors are solved
  exactly as a reference study for novel stochastic modeling and
  experimental setup.
\end{abstract}

\submitto{\NJP}
\maketitle

\section{Introduction}
Occurrence of Brownian yet non-Gaussian diffusion is increasingly
reported in experiments analyzing thermally driven motion in complex
biological contexts. Some examples are: (i) beads diffusing on lipid
tubes~\cite{wang2009}, in networks~\cite{wang2012,toyota2011}, or in a
matrix of micropillars~\cite{chakraborty2020}; (ii) tracers in
colloidal, polymeric and active
suspensions~\cite{weeks2000,wagner2017}; (iii) lipid molecules or
proteins embedded in protein-crowded lipid
membranes~\cite{jeon2016,yamamoto2017}, in biological
cells~\cite{stylianidou2014,parry2014,munder2016,cherstvy2018}, in
narrow corrugated channels with fluctuating
cross-section~\cite{li2019}, in anisotropic liquid
crystals~\cite{cuetos2018}; (iv) motion of individuals in
heterogeneous populations such as nematodes~\cite{hapca2008}; (v)
colloids in random force fields~\cite{pastore2021rapid}.  This
interesting phenomenon motivates mesoscopic modeling
invoking superposition of
statistics~\cite{beck2003,beck2006,hapca2008,wang2012}, diffusing
diffusivities~\cite{chubynsky2014,chechkin2017,jain2017, tyagi2017,
  miyaguchi2017, sposini2018, sposini2018first,miotto2021length}, 
subordination concepts~\cite{chechkin2017}, continuous time random
walk~\cite{barkai2020}, and diffusion in disordered
environments~\cite{sokolov2021},
but also calls for microscopic
foundation, as the identification of underlying universal mechanisms
could assist the interpretation of experimental evidence or inspire
new protocols.  In order to fill the latter gap, we recently
proposed~\cite{baldovin2019} that monomers aggregation and
disaggregation in the polymerization/depolymerization process offers a
natural foundation to such anomalous diffusion if one is monitoring
the motion of the center of mass (CM), an idea which has been further
explored in Ref.~\cite{hidalgo2020} through a many-body approach.
Profiting of this intuition, here we put forward a most transparent
universal phenomenon triggering Brownian yet non-Gaussian diffusion: a
polymer in contact with a chemostatted monomer bath.  By changing the
monomer concentration in the bath, the polymer undergoes a shift from
finite to infinite average growth, a contingency which separates the
dilute from the dense polymer
phase~\cite{deGennes1972,deGennes1979,vanderzande1998,madras2013}.  At
the transition point, diverging size fluctuations trigger an initial
non-Gaussian diffusion of the polymer CM which finally crosses over to
an ordinary diffusion dynamics.  We address this intriguing mechanism
in the simplest possible contest in which \textit{chain
  polymerization}~\cite{odian2004} occurs as a birth-death process,
and discuss it in the mean-field limit.  By establishing contact
with queuing theory, this enables us to obtain an explicit solution
of the model in which the (short-time) kurtosis of the polymer CM,
measuring the degree of non-Gaussianity, diverges as the
polymerization process becomes critical.  We emphasize that the
solution is general enough to deal with the several polymer models
known in the literature~\cite{deGennes1979,Doi1992} (Rouse, Zimm,
reptation), each giving rise to a different leptokurtic probability
density functions (PDF).  Moreover all basic, time-dependent
statistical quantities can be explicitly derived.
%\textbf{While former indications~\cite{baldovin2019} only relied on numerical
%  results, here we provide clear-cut analytical
%  evidence  in which all time-dependent statistical quantities
%  are explicitly derived. This brings to light the role of the
%  (previously hidden) polymerization critical point in rousing the
%  anomalous behavior.} 
We finally note that, while the crossover-time to Gaussian diffusion
is affected by critical slowing down, the reaction rate of the 
polymerization process still offers an independent parameter
controlling how fast normal diffusion is restored.   

The paper is organized as follows. In Section~\ref{sec_critical} we
introduce the grand
canonical partition function of chemostatted polymers at equilibrium
and the mean-field master equations describing the stochastic
polymerization/depolymerization process whose exact solution is
detailed in~\ref{sec_polymerization}. In Section~\ref{sec_bng} we study the stochastic
motion of the CM of chemostatted polymers and highlight its non-Gaussian behavior by
computing exactly the time evolution of the kurtosis under both equilibrium and
nonequilibrium conditions, together with the shape of the initial
non-Gaussian PDF under
equilibrium condition. These exact results are compared with 
Gillespie-Langevin simulations in~\ref{sec_gillespie}.
A summary and discussion is provided in
Section~\ref{sec_conclusions},
while further mathematical details are deferred to the appendixes.

\section{Critical polymers}
\label{sec_critical}
Chemostatted polymers are conveniently described in the grand canonical
ensemble where the monomer fugacity $z$  
%\textcolor{red}{
governs  the grand canonical partition function
  $Z_{\mathrm{gc}}$
  and the equilibrium distribution $P^\star_N(n)$
  associated to the event $N=n$ for
  the fluctuating polymer
  size $N$.
  Close to criticality,
  \mbox{$z\to z_{\mathrm{c}}^-$}, the former
%}
behaves asymptotically as~\cite{deGennes1972,deGennes1979,vanderzande1998,madras2013} 
\begin{equation}
  Z_{\mathrm{gc}}(z)=\sum_{n}(\mu_{\mathrm{c}}\,z)^n\;n^{\gamma-1}\,,
\end{equation}
where $\mu_{\mathrm{c}}$ is the (model-dependent) connective constant
and $z_{\mathrm{c}}=\mu_{\mathrm{c}}^{-1}$.  The universal entropic
exponent $\gamma$ is specified by the space dimension $d$, by the
underlying topology of the polymeric structure, and by the equilibrium phase:
good, $\Theta$-, or bad solvent (see,
e.g.,~\cite{nampoothiri2021lett} and references therein).  
The critical point $z=z_{\mathrm{c}}$ separates the dilute phase,
characterized by a finite average size $\mathbb{E}[N]$, from the dense
one in which the average size
diverges.

Interestingly, the equilibrium distribution
\begin{equation}
  P^\star_N(n)=\dfrac{(z/z_{\mathrm{c}})^n\;n^{\gamma-1}}{Z_{\mathrm{gc}}(z)}
\end{equation}
can be
related to a simple master equation describing the
polymerization/depolymerization process occurring as monomers add and
detach to the polymer in the grand canonical ensemble;
in this way connection with queing theory is established.
For convenience, if $n_{\mathrm{min}}$ is the minimal
polymer size in the chain polymerization process, we operate the
change of variable~\footnote{
Because of specificities of the mean-field limit we prefer this change of variable
with respect to the one adopted in~\cite{nampoothiri2021lett}. 
} 
$n\mapsto n-n_{\mathrm{min}}$ which associates to $P^\star_N(n)$ the support
$0\leq n<\infty$ without altering the asymptotic singular behavior close to criticality.
Consider then the (forward) master equation 
\begin{equation}
  \begin{array}{ll}
    \partial_t P_N(n,t|n_0)
    &=
    \mu\,P_N(n+1,t|n_0)
    +\lambda(n-1)\,P_N(n-1,t|n_0)
    \\
    &\quad
  -(\mu+\lambda(n))P_N(n,t|n_0)
    \quad\;\;(n>0)
    \vspace{0.1cm}
    \\
    \partial_t P_N(0,t|n_0)
    &=
    \mu\,P_N(1,t|n_0)
    -\lambda(0)\,P_N(0,t|n_0)\,.
  \end{array}
  \label{eq_master}
\end{equation}
Here $P_N(n,t|n_0)$ is the probability for $N=n$ at time $t\geq0$ given
$N=n_0$ at $t=0$, and $\lambda(n)$, $\mu$ are
the rates for association and dissociation, respectively.
In Eq.~\eqref{eq_master} it is assumed size-dependency for the association rate
(as it typically relies on the local concentration of available
monomers), whereas dissociation normally occurs independently of $n$.
Defining the growth factor as $g(n)\equiv \lambda(n)/\mu$,
%\textcolor{red}{
  in the Appendix we show
%}
that stationarity is attained under detailed
balance, $g(n)=P^\star_N(n+1)/P^\star_N(n)$: this identifies the
polymerization process, given $P^\star_N(n)$. Note that the rate $\mu$
remains a free parameter which may rescale Eq.~\eqref{eq_master},
thus determining the time scale $\tau$ for 
the autocorrelation of $N(t)$ (see below).

Knowledge of $\gamma$ is
provided by a mapping to the magnetic $O(n\to0)$
model~\cite{deGennes1979}; in the present paper we focus on the
mean-field
limit, in which $\gamma\to1$. This provides the simplification
\begin{equation}
  \lambda(n)=\lambda\,,\quad g(n)=g=z/z_{\mathrm{c}}\quad\forall n\,.
\end{equation}
In practical terms, the simplest situation we may associate to the
mean-field limit is that of a linear polymer composed by $N=n$
subunits $A_N$, subject to the chemical reaction~\cite{boal2002,odian2004}
$A_{N}+A_1\xrightleftharpoons[k_-=\mu]{k_+} A_{N+1}$,
where $k_+$, $k_-$ are the rate constants for association and dissociation,
respectively, and $\lambda=k_+\,c$ with $c$ the ($n$-independent)
local concentration of monomers $A_1$.
In this case, the grand canonical partition function and the equilibrium distribution
recast into 
\begin{equation}
 Z_{\mathrm{gc}}=\dfrac{1}{1-g}\,,
  \quad P_N^\star(n)=\mathrm{U}(n)\,(1-g)\,g^n=\mathrm{U}(n)\,(1-g)\,\mathrm{e}^{-n/\overline{n}} \,,
\end{equation}
with $\overline{n}=-1/\ln g$ and $\mathrm{U}(n)$ the (discrete) unit
step function.  As $g\to1^-$, the exponential distribution widens
tending to become uniform (over an infinite support). Consistently
with what anticipated, we thus appreciate that $g=1$ separates a $g>1$
phase of infinite growth from a $g<1$ phase of finite average size and
variance
\begin{equation}
  \mu_N\equiv\mathbb{E}[N]=\dfrac{g}{1-g}\,,
  \quad\sigma^2_N\equiv\mathbb{E}[N^2]-(\mathbb{E}[N])^2=\dfrac{g}{(1-g)^2}\,,
\end{equation}
respectively.
%\textcolor{red}{
  Observe that for $g\ll1$ one may write
  $\sigma^2_N=\mu_N/(1-g)\sim\mu_N$
  since $1-g$ is finite and $g$
  infinitesimal. On the other hand, as $g\to1^-$, the variance
  crosses over to the behavior
  $\sigma^2_N=\mu_N^2/g\sim\mu_N^2$; the latter being a signature of
  non-normal fluctuations arising at the critical point.
%}

The mean-field master equation can be written as
\begin{equation}
  \label{eq_poly}
  \begin{array}{ll}
  \partial_t P_N(n,t|n_0)
  &=
  \mu\,P_N(n+1,t|n_0)
  +g\,\mu\,P_N(n-1,t|n_0)
  \\
  &\quad
  -(g\,\mu+\mu)P_N(n,t|n_0)\,
  \qquad(n>0)
  \vspace{0.3cm}
  \\
  \partial_t P_N(0,t|n_0)
  &=
  \mu\,P_N(1,t|n_0)
  -g\,\mu\,P_N(0,t|n_0)\,,
  \end{array}
\end{equation}
and we see that in this case the polymerization process corresponds in
fact  to the $M/M/1$ model (Markovian interarrival
times/Markovian service times/1 server) in queuing
theory~\cite{Jain2007}.
The time-dependent solution of Eq.~\eqref{eq_poly} is recapitulated
in~\ref{sec_polymerization}, where it is also reported
the time scale $\tau$ of the exponential decay of
the auto-correlation coefficient,
$\rho(t)\simeq\mathrm{e}^{-t/\tau}$:
\begin{equation}
\tau=\dfrac{1+g}{(1-g)^2\,\mu}\,.
\end{equation}
Note that rescaling time by $\tau$, renders the model independent of the reaction
rate $\mu$.
The asymptotic behavior for small and large time of $P_N(n,t|n_0)$ is
\mbox{$P_N(n,t|n_0)\substack{\sim\\t\ll\tau}\delta_{n,n_0}$},
\mbox{$P_N(n,t|n_0)\substack{\sim\\t\gg\tau}P^\star_N(n)$},
respectively.

\section{Brownian non-Gaussian diffusion}
\label{sec_bng}
A basic idea to provide diffusing diffusivity models with a
microscopic footing is very simple. From polymer physics it is known
that the CM position $\boldsymbol{R}_{\mathrm{CM}}$ of a 
macromolecule with $N+n_{\mathrm{min}}$ subunits diffuses with a
coefficient $D(N)=D_0/(N+n_{\mathrm{min}})^\alpha$, $D_0$ being
specific to the subunit, and $\alpha$ to the chosen polymer
model~\cite{deGennes1979,Doi1992}.
%\textcolor{red}{
  Notably, if the chain undergoes polymerization then
  $N=N(t)$ becomes a stochastic process, and so
  does the diffusion coefficient $D$.
%}
%\textcolor{red}{
  In slow nucleation processes $n_{\mathrm{min}}$ corresponds to the
  size of a polymer nucleus~\cite{oosawa1970}, and to keep contact with
  the exactly-solvable $M/M/1$ model the association and dissociation
  rates must be size-independent. By focusing on  a
  linear polymer which can grow and deteriorate at both ends, 
  in the following we consider $n_{\mathrm{min}}$=3. This means to assume
  the trimer as the minimal polymer conformation in the system.
  In this way,
  monomer can always attach and detach at the two extremities and
  $\lambda$, $\mu$ do not depend on the polymer size.
%}
Zimm model includes hydrodynamic interactions
and via the Stokes-Einstein relation~\cite{Doi1992,metzler2021} the
diffusion coefficient is proportional to the inverse of the polymer
hydrodynamic radius $R$, $D(N)\sim 1/R(N)\sim1/N^\nu$; hence, $\alpha$
coincides with the mean-field metric exponent $\nu$, $\alpha=\nu=1/2$.
Rouse dynamics is instead characterized by $\alpha=1$, and for
reptation $\alpha=2$.
On the Smoluchowski time
scale~\footnote{
The Smoluchowski time scale is appropriate for
the description of the polymer dynamics at any sizes
$N$ varying from single monomer to large
colloids. Indeed, in the time needed to loose
memory of inertial effects, the traveled distance
with respect to the polymer radius $a$ is given by
$$\displaystyle\dfrac{\sqrt{3\,m\,k_{\mathrm{B}}T}}{6\pi\,\eta\,a^2}\,.$$
In water at room temperature, this ratio varies from
$10^{-3}$ for single 
nucleotides or amino acids to $10^{-5}$ for large
colloids.
}, $\boldsymbol{R}_{\mathrm{CM}}$ evolves
according to
\begin{equation}
  \mathrm{d}\boldsymbol{R}_{\mathrm{CM}}(t)=\sqrt{2D(N(t))}
  \,\mathrm{d}\boldsymbol{B}(\mathrm{d}t)\,,
\label{eq_lang_1}
\end{equation}
where
%\textcolor{red}{
  $\boldsymbol{B}(t)=\mathcal{N}(0,t)$ is a Wiener process (Brownian
  motion) with infinitesimal increments
  $\mathrm{d}\boldsymbol{B}(\mathrm{d}t)=\mathcal{N}(0,\mathrm{d}t)$
  -- The notation $\mathcal{N}(\mu,\sigma^2)$ indicates a Gaussian random
  variable with average $\mu$ and variance $\sigma^2$.
%}
The same behavior is observed for any tagged monomer above the Rouse
relaxation time~\cite{deGennes1979,Doi1992}.
%\textcolor{red}{
  It is important to stress that in Eq.~\eqref{eq_lang_1} two sources
  of randomness are present. One is the standard thermal agitation
  imparted by the solvent and represented by $\boldsymbol{B}(t)$, the other is
  the polymerization process which affects the intensity of
  $\boldsymbol{B}(t)$ through the $N$-dependence of the diffusion
  coefficient. Technically, $\boldsymbol{B}(t)$ is referred to as the
  ``subordinated process'' and $N(t)$ as the ``subordinator
  process''. 
%}
Under ordinary conditions, the stationary distribution of $N$ is strongly
peaked around its mean value and the ``diffusion of diffusivities effect''
is then difficult to detect. The situation drastically changes in proximity 
of the critical point governing the divergence of the polymerization degree. 

\begin{figure}[t]
  \begin{center}
    \includegraphics[width=0.5\columnwidth]{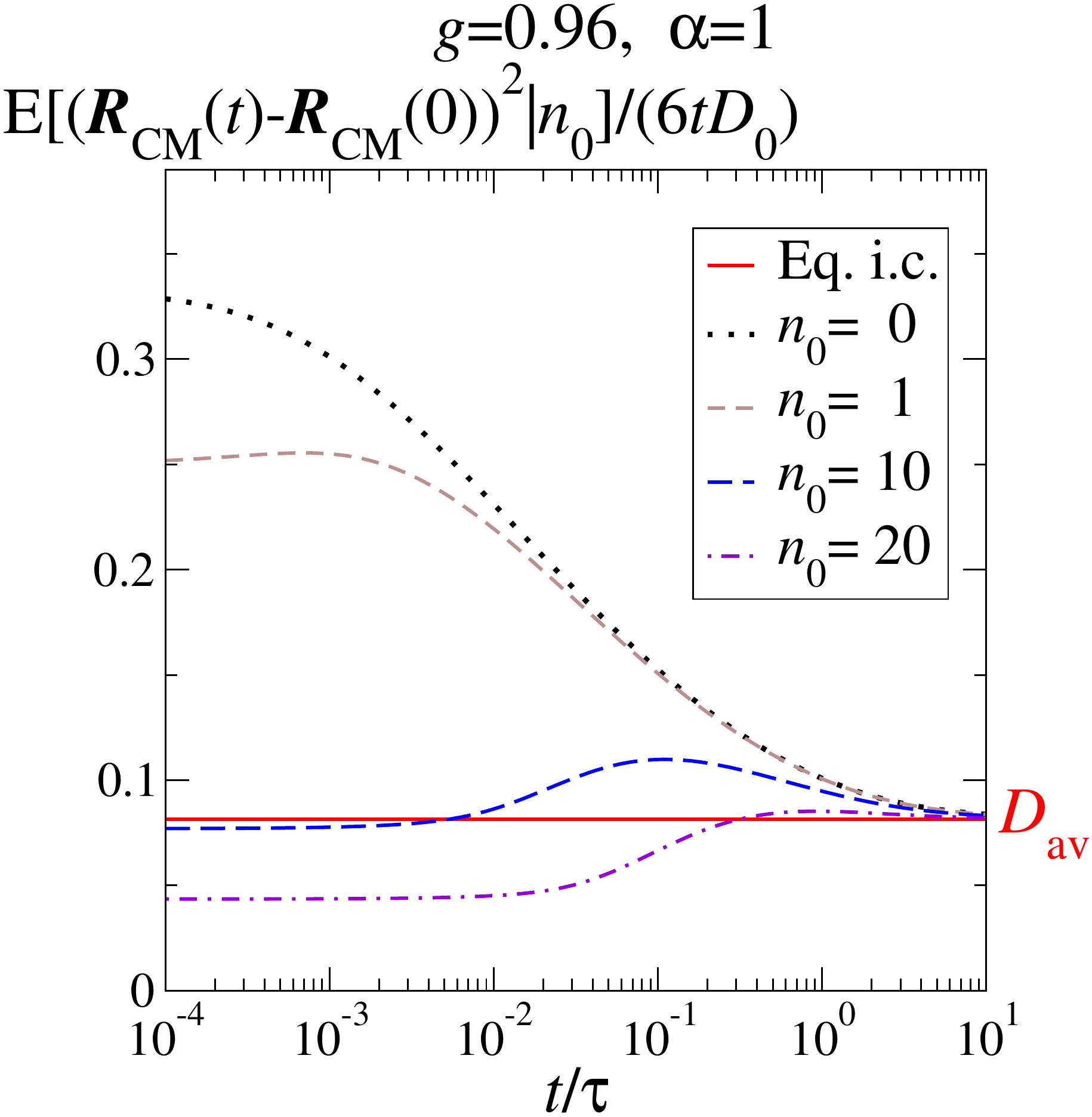}\\
  \end{center}
  \caption{Time evolution of the CM second moment with nonequilibrium
    initial conditions (dashed lines). Equilibrium initial
    conditions, $P_N(n_0,0)=P^\star_N(n_0)$ (full line) emphasize the
    Brownian character of the anomalous diffusion. Here and in the following
    plots, the analytical results have been confirmed by
    Gillespie simulations~\cite{gillespie1977} of the stochastic
    processes (see Appendix).}
  \label{fig_second_moment}
\end{figure}

%\textcolor{red}{
  For any given realization
  $[n(t)]\equiv\{n(t')\in\mathbb{N}\;|\;0\leq t'\leq t\}$ of the
  stochastic process $N(t)$, the PDF of the CM location satisfies an
  ordinary diffusion equation:
%}
\begin{equation}
  \partial_t p_{\boldsymbol{R}_{\mathrm{CM}}}(\boldsymbol{r},t|[n(t)];\boldsymbol{r}_0)
  =\dfrac{D_0}{(n(t)+3)^\alpha}\boldsymbol{\nabla}^2
  p_{\boldsymbol{R}_{\mathrm{CM}}}(\boldsymbol{r},t|[n(t)];\boldsymbol{r}_0)\,,
  \label{eq_diff_1}
\end{equation}
where $\boldsymbol{r}_0$ is the initial CM position.
%\textcolor{red}{
  Eq.~\eqref{eq_diff_1} emphasizes the fact that, to determine 
  $p_{\boldsymbol{R}_{\mathrm{CM}}}(\boldsymbol{r},t|[n(t)];\boldsymbol{r}_0)$
  at a given time $t>0$, the knowledge of the whole history of the
  subordinator process is required. Only in this way at each 
  update $\mathrm{d}t$ the correct diffusion coefficient  can be provided to
  propagate the initial condition
  $p_{\boldsymbol{R}_{\mathrm{CM}}}(\boldsymbol{r},0|n_0;\boldsymbol{r}_0)
  =\delta(\boldsymbol{r}-\boldsymbol{r}_0)$
  up to time $t$.
%}
%\textcolor{red}{
By  exploiting the scaling property
  $c\,\mathcal{N}(\mu,\sigma^2)=\mathcal{N}(c\,\mu,c^2\,\sigma^2)$ for
  $c\in\mathbb{R}$,  
%}
the diffusing path is conveniently reparametrized in terms of a
coordinate which converts Eq.~\eqref{eq_lang_1} into a standard overdamped
Langevin equation
%\textcolor{red}{
  with unit diffusion coefficient,
%} 
\begin{equation}
  \mathrm{d}\boldsymbol{R}_{\mathrm{CM}}(s)=
  \mathrm{d}\boldsymbol{B}(\mathrm{d}s)\,,
  \quad
%  \textcolor{red}{
    \mathrm{d}s=2\,D(n(t))\,\mathrm{d}t\,,
%  }
\end{equation}
where $s\geq0$ is a path variable corresponding to the realization of 
the stochastic process
\begin{equation}
  S(t)\equiv\int_0^t\mathrm{d}t'\,2\,D(N(t'))\,.
  \label{eq_s_t}
\end{equation}
%\textcolor{red}{
  With respect to the ``random path'' $s$, Eq.~\eqref{eq_diff_1}
  formally transforms into an ordinary diffusion equation~\cite{risken1996},
%}
\begin{equation}
  \partial_s p_{\boldsymbol{R}_{\mathrm{CM}}}(\boldsymbol{r},s|\boldsymbol{r}_0)
  =\boldsymbol{\nabla}^2
  p_{\boldsymbol{R}_{\mathrm{CM}}}(\boldsymbol{r},s|\boldsymbol{r}_0)\,,
  \label{eq_diff_2}
\end{equation}
with Green function solution
\begin{equation}
  p_{\boldsymbol{R}_{\mathrm{CM}}}(\boldsymbol{r},s|\boldsymbol{r}_0)
  =\dfrac{1}{\left(2\pi\,s\right)^{3/2}}
  \,\exp\left(
  -\dfrac{
    (\boldsymbol{r}-\boldsymbol{r}_0)^2
  }{
    2\,s}
  \right)\,.
\end{equation}
For a polymer of size $n_0+3$ starting with certainty at the origin,
$p_{\boldsymbol{R}_{\mathrm{CM}}}(\boldsymbol{r},0|n_0;\boldsymbol{0})
=\delta_{n,n_0}\,\delta(\boldsymbol{r})$,
the PDF of finding its CM at position $\boldsymbol{r}$ at time $t$ is
thus given by the subordination~\cite{Feller1968,bochner2020harmonic} formula
\begin{equation}
  p_{\boldsymbol{R}_{\mathrm{CM}}}(\boldsymbol{r},t|n_0;\boldsymbol{0})
  =\int_0^\infty\mathrm{d}s
  \,\dfrac{\mathrm{e}^{-\frac{\boldsymbol{r}^2}{2s}}}{(2\pi\,s)^{3/2}}
  %p_{\boldsymbol{R}_{\mathrm{CM}}}(\boldsymbol{r},s|\boldsymbol{0})
  \;p_S(s,t|n_0)\,,
  \label{eq_subordination}
\end{equation}
where $p_S(s,t|n_0)$ is the probability distribution of the process
$S(t)$. Eq.~\eqref{eq_subordination} makes explicit the
non-Gaussianity of the CM diffusion: while a broad
distribution for $S(t)$ implies a fat-tailed Gaussian mixture PDF 
$p_{\boldsymbol{R}_{\mathrm{CM}}}(\boldsymbol{r},t|n_0,\boldsymbol{0})$, when
$p_S(s,t|n_0)$ concentrates around a specific value, the normal, Gaussian
behavior of the CM diffusion is restored. 

From Eq.~\eqref{eq_subordination} we also have
\begin{equation}
  \mathbb{E}[(\boldsymbol{R}_{\mathrm{CM}}(t)-\boldsymbol{R}_{\mathrm{CM}}(0))^2|n_0]
  =3\,\mathbb{E}[S(t)|n_0]
  =3\sum_{n=0}^\infty\dfrac{2\,D_0}{(n+3)^\alpha}\,
  \int_0^t\mathrm{d}t'P_N(n,t'|n_0).
  \label{eq_av_s}
\end{equation}
The short time expansion of $P_N(n,t|n_0)$ reported in the Appendix
exhibits nonlinear corrections (alternating series) to the
linear increase of the mean squared displacement.
The Brownian character becomes distinctive of large time, $t\gg\tau$, when
$P_N(n,t|n_0)\simeq P^\star_N(n)$ and Eq.~\eqref{eq_av_s} simplifies to: 
\begin{equation}
  \mathbb{E}[(\boldsymbol{R}_{\mathrm{CM}}(t)-\boldsymbol{R}_{\mathrm{CM}}(0))^2|n_0]
  =6\,D_{\mathrm{av}}\,t\quad(t\gg\tau)\,,
  \label{eq_brownian_2}
\end{equation}
with $D_{\mathrm{av}}\equiv\displaystyle\sum_{n=0}^\infty \dfrac{D_0}{(n+3)^\alpha}\,P^\star_N(n)$.
In a potential experimental protocol in which the CM
diffusion is monitored by starting from a  polymer of a given  size $\overline{n}_0+3$, 
nonlinear corrections in time mark the crossover to
the final Brownian regime in Eq.~\eqref{eq_brownian_2}.
At variance, in an arrangement in which the initial polymer sizes are
distributed according to equilibrium, $P_N(n_0,0)=P^\star_N(n_0)$,
Eq.~\eqref{eq_brownian_2} turns out to be valid at all time $t\geq0$.
Fig.~\ref{fig_second_moment} summarizes these behaviors.

It is common practice to quantify the non-Gaussian behavior in terms of
the kurtosis
\begin{equation}
  \kappa_X(t|n_0)\equiv\dfrac{
    \mathbb{E}\left[\left(X_{\mathrm{CM}}(t)-\mathbb{E}[X_{\mathrm{CM}}(t)]\right)^4|n_0\right]
  }{
   \left(\mathbb{E}\left[\left(X_{\mathrm{CM}}(t)-\mathbb{E}[X_{\mathrm{CM}}(t)]\right)^2|n_0\right]\right)^2
  }
\end{equation}
($\kappa_X=3$ for Gaussian variables).
Note that for simplicity here we refer only to the $x$-component of the
random vector $\boldsymbol{R}_{\mathrm{CM}}\equiv(X_{\mathrm{CM}},Y_{\mathrm{CM}},Z_{\mathrm{CM}})$.
From the subordination formula, Eq~\eqref{eq_subordination}, we get
\begin{equation}
  \kappa_X(t|n_0)=3\dfrac{\mathbb{E}[S^2(t)|n_0]}{\left(\mathbb{E}[S(t)|n_0]\right)^2}\,,
  \label{eq_kurtosis}
\end{equation}
and using again Eq.~\eqref{eq_s_t},
\begin{equation}
  \mathbb{E}[S^2(t)|n_0]
  =2\sum_{n',n''=0}^\infty
  \dfrac{2\,D_0}{(n'+3)^\alpha}
  \,\dfrac{2\,D_0}{(n''+3)^\alpha}
  \int_0^t\mathrm{d}t''
  \int_0^{t''}\mathrm{d}t'
  P_N(n'',t''|n',t')
  \;P_N(n',t'|n_0)\,,
  \label{eq_av_s_2}
\end{equation}
where we have used the Markov property
$P_N(n'',t''|n',t',n_0)=P_N(n'',t''|n',t')$ for $t''\geq t'\geq0$.
Integrals in Eqs.~\eqref{eq_av_s},~\eqref{eq_av_s_2}
can be performed, e.g., through Eq.~\eqref{eq_morse_solution} in the
Appendix, so that the function  $\kappa_X(t|n_0)$ can be calculated exactly.

\begin{figure}[t]
  \begin{center}
    \includegraphics[width=0.5\columnwidth]{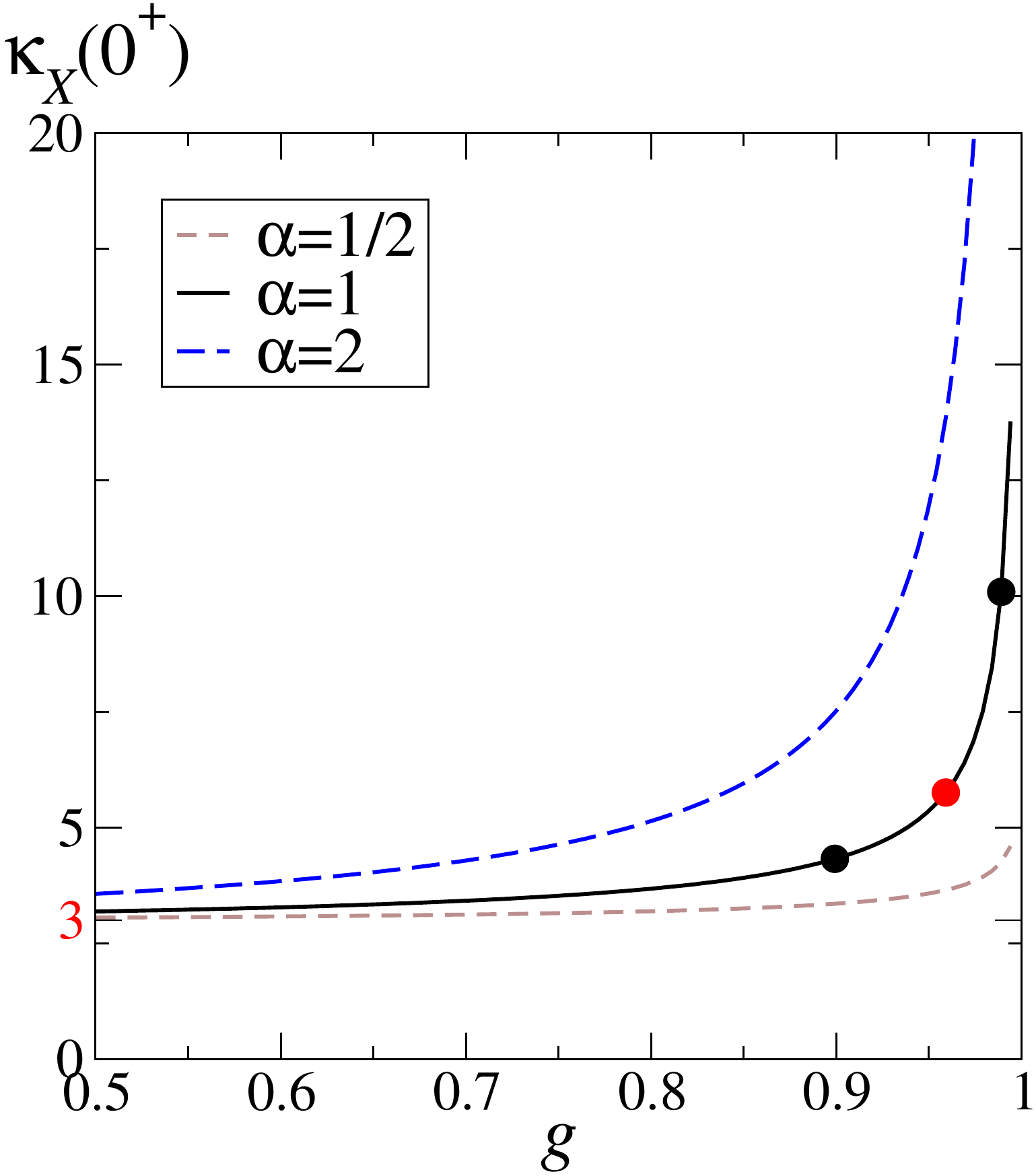}\\
  \end{center}
  \caption{Initial kurtosis when starting with equilibrium polymer
    sizes. For all polymer models the kurtosis diverges at the critical
    value $g=1$. Full circles indicate values for the PDFs reported in
    Fig.~\ref{fig_initial_pdf}, with the red one ($g=0.96$) also visible
    in Fig.~\ref{fig_kurtosis_evolution}.} 
  \label{fig_kurtosis_initial}
\end{figure}
\begin{figure}[t]
  \begin{center}
    \includegraphics[width=0.5\columnwidth]{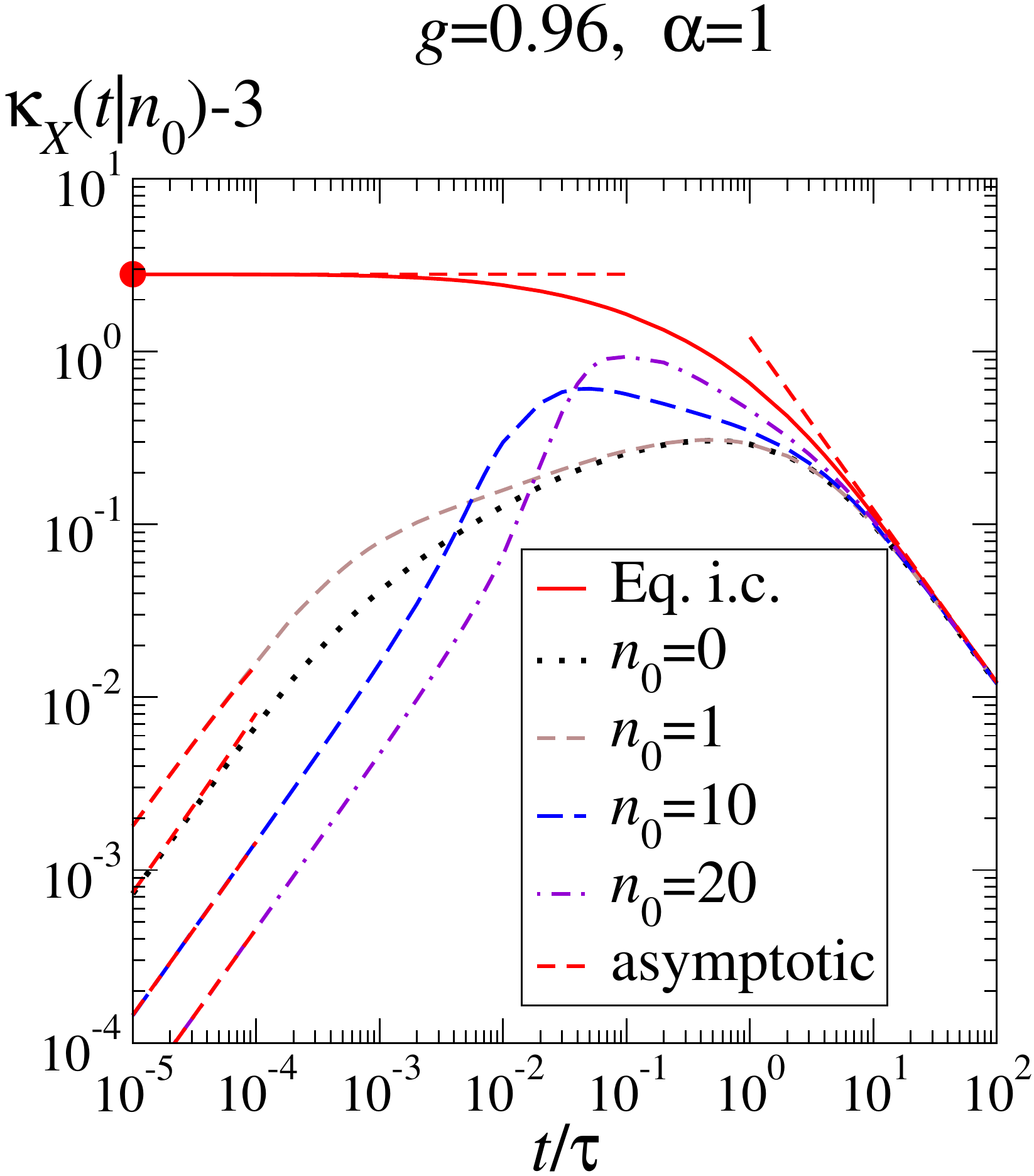}\\
  \end{center}
  \caption{Time evolution of the CM $x$-kurtosis with nonequilibrium
    (dashed lines) and equilibrium initial
    conditions (full line). Red dashed lines highlight asymptotic predictions
    obtained through
    Eqs.~\eqref{eq_kurtosis},~\eqref{eq_av_s_2},~\eqref{eq_morse_solution},~\eqref{eq_short_time_expansion}. 
    The red circle can be retraced in Fig~\ref{fig_kurtosis_initial} and
    the related PDF is visible in Fig.~\ref{fig_initial_pdf}.
  }  
  \label{fig_kurtosis_evolution}
\end{figure}

Summarizing,  we can depict two different scenarios for the CM
dynamics of a polymer undergoing polymerization/depolymerization when
in contact with a chemostatted monomer bath: For a protocol with an initial specified
polymer size $n_0+n_{\mathrm{min}}$
($P_N(n_0,0)=\delta_{n_0,\overline{n}_0}$),
Eqs.~\eqref{eq_kurtosis},~\eqref{eq_av_s_2},~\eqref{eq_short_time_expansion}
imply an initial Gaussian behavior with $\kappa_X(0^+|n_0)=3$. As the distribution in polymer sizes spreads,
$\kappa_X$ grows, reaches a maximum, and then returns to $\kappa_X(t|n_0)\sim3$
for $t\gg\tau$ as specified below.
If one starts instead with the  equilibrium initial conditions $P_N(n_0,0)=P^\star_N(n_0)$,
Eq.~\eqref{eq_av_s_2} simplifies to
\begin{equation}
  \mathbb{E}[S^2(t)]
  =2\sum_{n',n''=0}^\infty
  \dfrac{2\,D_0}{(n'+3)^\alpha}
  \,\dfrac{2\,D_0}{(n''+3)^\alpha}
  \;P^\star_N(n')
  \int_0^t\mathrm{d}t''
  \int_0^{t''}\mathrm{d}t'
  P_N(n'',t''|n',t')\,,
  \label{eq_av_s_2_eq}
\end{equation}
which together with
Eqs.~\eqref{eq_kurtosis},~\eqref{eq_short_time_expansion}
provide the initial kurtosis
\begin{equation}
  \kappa_X(0^+)=3\dfrac{
    \mathbb{E}\left[(N(0)+3)^{-2\alpha}\right]
  }{
    \left(\mathbb{E}\left[(N(0)+3)^{-\alpha}\right]\right)^2
  }
  =3\dfrac{\Phi(g,2\alpha,3)}{(1-g)\,\Phi^2(g,\alpha,3)}\,,
  \label{eq_kurtosis_initial}
\end{equation}
where $\Phi(g,\alpha,3)\equiv\sum_{n=0}^\infty g^n/(n+3)^\alpha$ is
the Lerch transcendent function \cite{Olver2010}.
%Also in this case for $t\gg\tau$ the
%asymptotic value $\kappa_X(t)\sim3$ is approached.
Fig.~\ref{fig_kurtosis_initial} plots Eq.~\eqref{eq_kurtosis_initial}
showing that at the critical point $g=1$ the
equilibrium initial 
kurtosis diverges for all models of polymer dynamics.
Note that for polymer reptation where $\alpha=2$,  we have a power law divergence
($\kappa_X(0^+)\sim(1-g)^{-1}$), which is
logarithmic corrected ($\kappa_X(0^+)\sim(1-g)^{-1}\,|\ln(1-g)|^{-2}$  
for the Rouse dynamics ($\alpha=1$).
In the Zimm  model ($\alpha=1/2$) the divergence is
logarithmic ($\kappa_X(0^+)\sim-\ln(1-g)$)\footnote{
%\textcolor{red}{
  For Zimm and Reptation dynamics, one should also consider
  intrinsic fluctuations of the diffusion coefficient due to
  long-range effects triggered by hydrodynamics and entanglement,
  respectively.  For equilibrium initial conditions
  such effects add a correction~\cite{miyaguchi2017} to
  the initial kurtosis which becomes negligible at the critical point
  $g\to1$, where the fluctuations of the polymer size due to the
  polymerization/degradation processes dominate the dynamics. On the
  contrary, for out-of-equilibrium initial condition the intrinsic
  fluctuations addressed in Ref.~\cite{miyaguchi2017} produce a
  small-time correction  to the plots in
  Fig.~\ref{fig_kurtosis_evolution}. 
%}
}.   
At large time $t\gg\tau$, the kurtosis does not depend on the 
initial conditions anymore; 
through Eq.~\eqref{eq_av_s_2_eq} and~\eqref{eq_morse_solution} it is
easy to prove  a universal power-law decay for the excess kurtosis,
independent of the polymer model:
$\kappa_X(t|n_0)-3\,\substack{\propto\\t\gg\tau}\,1/t$.
In Fig.~\ref{fig_kurtosis_evolution} we report the evolution of the kurtosis,
for both equilibrium and nonequilibrium initial polymer sizes. 
%\textcolor{red}{
  We observe that close to the critical point, where $P_N^\star(n)$
  tends to be uniform over an infinite support,
  the equilibrium results become largely independent of the specific
  choice of $n_{\mathrm{min}}$. 
  On the contrary, the value of $n_{\mathrm{min}}$ affects the
  nonequilibrium behavior e.g. in situations where a specific initial
  size $n_0$ is considered.
%}

\begin{figure}[t]
  \begin{center}
    \includegraphics[width=0.5\columnwidth]{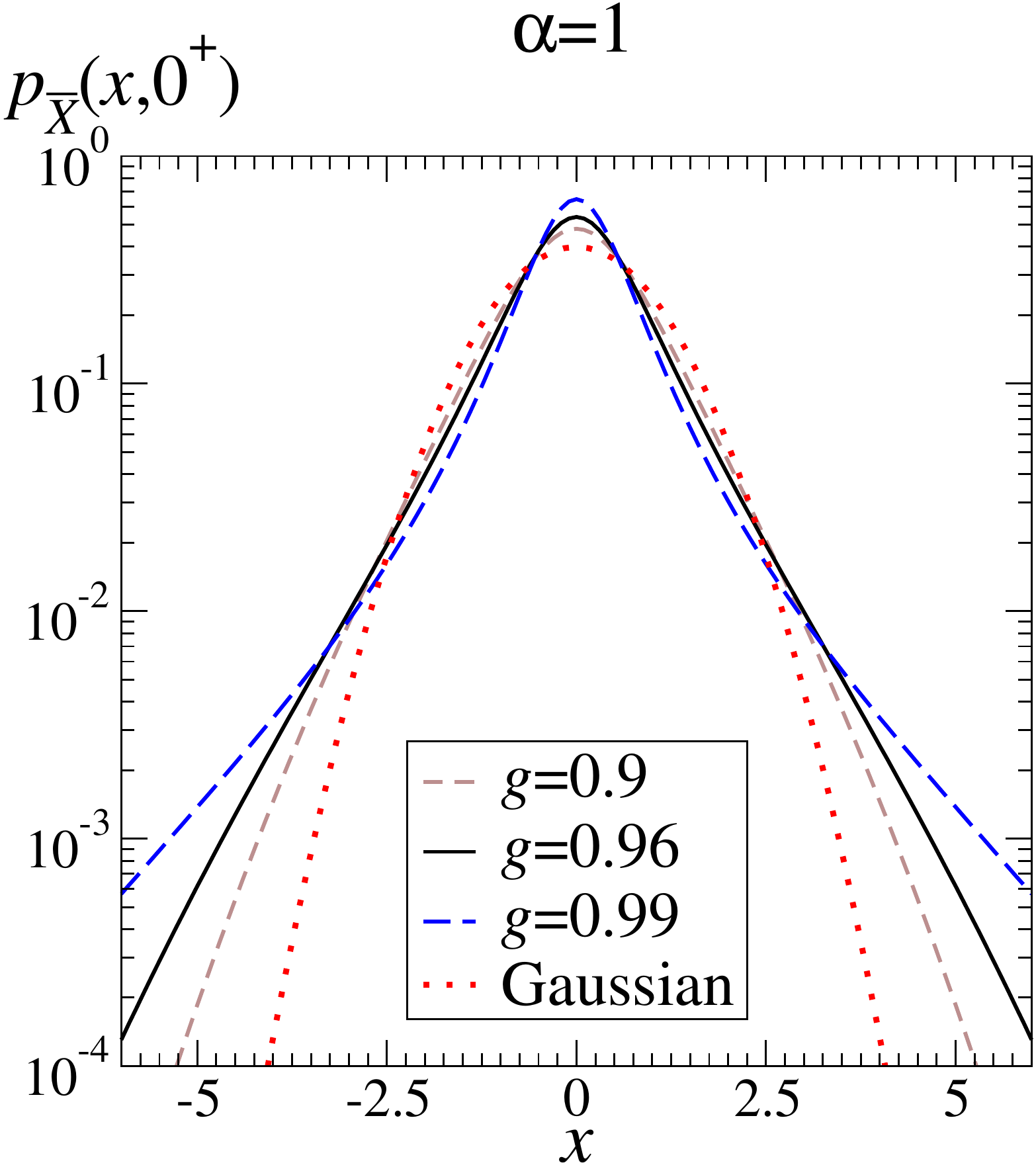}\\
  \end{center}
  \caption{Unit-variance initial $x$-PDF for the CM of a Rouse polymer
    with equilibrium sizes $P_N(n_0,0)=P^\star_N(n_0)$.
    For comparison purposes, a unit-variance Gaussian
    PDF is also plotted in red dotted line.}
  \label{fig_initial_pdf}
\end{figure}

Under equilibrium conditions, it is interesting to look at the
shape of the initial non-Gaussian PDF for the polymer CM.
In order to do so, it is convenient to switch to the unit-variance
dimensionless variable
$\overline{X}_{\mathrm{CM}}(t)\equiv X_{\mathrm{CM}}(t)/\sqrt{\mathbb{E}[X^2(t)]}$.
From Eq.~\eqref{eq_subordination}, as $t\to0^+$, we have
\begin{equation}
  p_{\overline{X}}(x,0^+)
  =\sum_{n=0}^\infty P^\star_N(n)
  \,\dfrac{\mathrm{e}^{-\frac{D_{\mathrm{av}}\,(n+3)^\alpha\,x^2}{2D_0}}
  }{
    \sqrt{2\pi\,\frac{D_0}{D_{\mathrm{av}}\,(n+3)^\alpha}}
  }\,,
\end{equation}
which only depends on $g$ and $\alpha$.
As displayed in Fig.~\ref{fig_initial_pdf}, the tails of this PDF
increase with  $g$. At large $|x|$ the PDF is
asymptotic to the Gaussian cutoff
$\sim\mathrm{e}^{-3^\alpha D_{\mathrm{av}}x^2/(2D_0)}$, and as $g\to1$
this cutoff is pushed towards $|x|\to\infty$.
Consistently with the divergent behavior of the initial kurtosis,
tails of the PDF similarly increase with $\alpha$ (see
Fig.~\ref{fig_initial_pdf_sm} in the Appendix).

\section{Conclusions}
\label{sec_conclusions}
We have proposed an analytically solvable microscopic model underpinning 
Brownian yet non-Gaussian diffusion, a phenomenon ubiquitous  
in many soft matter and biological  systems. 
The model describes the  diffusion dynamics of 
the CM of a polymerizing/depolymerizing chain
in contact with a chemostatted monomer bath.
%\textcolor{red}{
  A situation of potential biological importance for such a scenario is that of
  polymers that bind at specific receptors on cell membranes. In
  this case the size fluctuations of the tagged particle not only
  strongly affect the diffusion coefficient of the polymer and thus its mean first
  passage time to the target, but, once bounded, it may also influence
  the efficiency of the polymer to trespass the membrane.    
%}
We have shown that monomer concentration emerges as a natural
experimental parameter to study the system under critical
conditions. Likewise critical opalescence, we have found that the
initial kurtosis $\kappa$ of the CM location becomes a power-law diverging
\textit{dynamical response} when the monomer concentration $c\to k_-/k_+$ ($g\to1$).
Another conceivable experimental parameter is  the concentration
of chemically inert macromolecules, added in 
solution. Starting at low concentrations,
by increasing the crowdness of the environment, it is possible to convert 
dynamical conditions from the Zimm-hydrodynamic
regime to Rouse-diffusion and up to reptation~\cite{deGennes1979,Doi1992}, 
thus effectively modifying $\alpha$. This is yet another possible
experimental route to
emphasize non-Gaussian behavior. 
Finally,
the established connection with queuing theory may reveal to be useful also
where more complex polymerization/depolymerization processes lead to
intricate polymer topologies like  branched polymers. Extensions
of the theory along these directions would be very interesting.  

\section*{Acknowledgments}
We acknowledge insightful discussions with R. Metzler. This work has
been partially supported by the University of Padova BIRD191017
project ``Topological statistical dynamics''.

%\section*{Appendix}
\appendix

%\textcolor{red}{
  \section{Proof of $g(n)=P^\star_N(n+1)/P^\star_N(n)$}
  Defining $g(n)\equiv\lambda(n)/\mu$,
  the stationary solution $P^\star_N(n)$ of Eq.~\eqref{eq_master} satisfies
\begin{equation}
  \begin{array}{ll}
    0
    &=
    P^\star_N(n+1)
    +g(n)\,P^\star_N(n-1)
    -(1+g(n))\,P^\star_N(n)
    \quad\;\;(n>0)
    \vspace{0.1cm}
    \\
    0
    &=
    P^\star_N(1)
    -g(0)\,P^\star_N(0)\,.
  \end{array}
\end{equation}
We thus have $g(0)=P^\star_N(1)/P^\star_N(0)$. By induction it then
follows \mbox{$g(n)=P^\star_N(n+1)/P^\star_N(n)\,.\quad\Box$}
%}

\section{Polymerization -- Queuing $M/M/1$ model}
\label{sec_polymerization}
%\subsection{Polymerization -- Queuing $M/M/1$ model}
We briefly review the
%\textcolor{red}{
  mean-field
%}
solution~\cite{Jain2007} for the $P_N(n,t|n_0)$
fulfilling
%\textcolor{red}{
\begin{equation}
  \begin{array}{ll}
    \partial_t P_N(n,t|n_0)
    &=
    \mu\,P_N(n+1,t|n_0)
    +\lambda\,P_N(n-1,t|n_0)
    \\
    &\quad
  -(\mu+\lambda)P_N(n,t|n_0)
    \quad\;\;(n>0)
    \vspace{0.1cm}
    \\
    \partial_t P_N(0,t|n_0)
    &=
    \mu\,P_N(1,t|n_0)
    -\lambda\,P_N(0,t|n_0)\,.
  \end{array}
    \label{eq_poly_app}
\end{equation}
%}
The probability generating
function
$\displaystyle G(z,t|n_0)\equiv\sum_{n=0}^\infty z^n\,P_N(n,t|n_0)$
satisfies
\begin{equation}
  \label{eq_generating}
  z\,\partial_t G_N(z,t|n_0)=
  (1-z)\,[(\mu-\lambda\,z)\,G_N(z,t|n_0)-\mu\,G_N(0,t|n_0)]\,,
\end{equation}
with stationary solution
\begin{equation}
  G_N^\star(z)=\dfrac{1-g}{1-g\,z}
  \;\Leftrightarrow
  P_N^\star(n)=\mathrm{U}(n)\,(1-g)\,g^n\,,
\end{equation}
$\mathrm{U}(n)$ being the (discrete) unit step function and $g\equiv\lambda/\mu$.
Applying the Laplace transform,
$\overline{(\cdot)}\equiv\int_0^\infty\mathrm{d}t\,\mathrm{e}^{-\theta\,t}(\cdot)$,
to Eq.~\eqref{eq_generating},  one gets
\begin{equation}
  \overline{G}_N(z,\theta|n_0)
  =\dfrac{
    z^{n_0+1}-\mu(1-z)\,\overline{P}_N(0,\theta|n_0)
  }{
    (\lambda+\mu+\theta)\,z-\mu-\lambda\,z^2\,.
  }
\end{equation}
Analyzing the zeroes $z_1(\theta)$, $z_2(\theta)$  of the denominator
of the last expression and using Rouch\'e's theorem leads to
\begin{equation}
  \overline{P}_N(0,\theta|n_0)
  =\dfrac{[z_1(\theta)]^{n_0+1}}{\mu[1-z_1(\theta)]}\,,
\end{equation}
where $z_1$ is the zero with $|z_1|<1$.
At this point $\overline{G}_N(z,\theta|n_0)$ can be explicitly written
as a series expansion, in order to  identify
$\overline{P}_N(n,\theta|n_0)$. Finally, application of the inverse
Laplace transform provides the desired solution:
\begin{eqnarray}
  P_N(n,t|n_0)
  &=&\mathrm{U}(n)\,\mathrm{U}(n_0)\,\mathrm{e}^{-(1+g)\,\mu t}
  \;
  \left[
    g^{\frac{n-n_0}{2}}
    \,I_{n-n_0}\left(2\sqrt{g}\,\mu t\right)
    \right.
    \nonumber\\
    &&\;\;\;\left.
    +g^{\frac{n-n_0-1}{2}}
    \,I_{n+n_0+1}\left(2\sqrt{g}\,\mu t\right)
    \right.
    \\
    &&\;\;\;\left.
    +(1-g)\,g^n
    \sum_{i=n+n_0+2}^\infty
    g^{-\frac{i}{2}}
    \,I_{i}\left(2\sqrt{g}\,\mu t\right)
    \right]\,
  \nonumber
\end{eqnarray}
where $I_n(z)$ is the modified Bessel function of the first
kind. Using the asymptotic behavior of the modified Bessel function,
it can be seen that $\lim_{t\to\infty}P_N(n,t|n_0)=P_N^\star(n)$.

An integral representation of this solution is given by~\cite{morse1955}
\begin{equation}
  \label{eq_morse_solution}
  \begin{array}{ll}
    P_N(n,t|n_0)
    &=\mathrm{U}(n)\,\delta_{n,n_0}
    -\mathrm{U}(n)\,\mathrm{U}(n_0)\,\dfrac{g^{\frac{n-n_0}{2}}}{\pi}
    \displaystyle\int_0^{2\pi}\mathrm{d}\theta
    \left[
      \sin(n_0\theta)-\sqrt{g}\sin((n_0+1)\theta)
      \right]\cdot
    \\
    &\qquad
    \cdot\,\left[
      \sin(n\theta)-\sqrt{g}\sin((n+1)\theta)
      \right]\,\dfrac{1-\mathrm{e}^{-[(1+g)-2\sqrt{g}\cos(\theta)]\,\mu t}
    }{
      \left[(1+g)-2\sqrt{g}\cos(\theta)\right]
    }\,,
  \end{array}
\end{equation}
%\textcolor{red}{
  where we point out the identity, obtained by taking the limit $t\to\infty$:
  \begin{equation}
    \label{eq_morse_identity}
    \begin{array}{ll}
      \mathrm{U}(n)\,(1-g)\,g^n
      &=\mathrm{U}(n)\,\delta_{n,n_0}
      -\mathrm{U}(n)\,\mathrm{U}(n_0)\,\dfrac{g^{\frac{n-n_0}{2}}}{\pi}
      \displaystyle\int_0^{2\pi}\mathrm{d}\theta
      \left[
        \sin(n_0\theta)-\sqrt{g}\sin((n_0+1)\theta)
        \right]\cdot
      \\
      &\qquad
      \cdot\,\left[
        \sin(n\theta)-\sqrt{g}\sin((n+1)\theta)
        \right]\,
      \dfrac{1
      }{
        \left[(1+g)-2\sqrt{g}\cos(\theta)\right]
      }\,.
    \end{array}
  \end{equation}
%}
From this expression, the auto-correlation coefficient
turns out to be
\begin{eqnarray}
  \rho(t)&\equiv&
  \dfrac{\mathbb{E}[N(t)\,N(0)]-(\mathbb{E}[N])^2}{\mathbb{E}[N^2]-(\mathbb{E}[N])^2}
  \nonumber\\
  &=&
  \dfrac{
    \displaystyle
    \sum_{n=0}^\infty\sum_{n_0=0}^\infty
    n\,n_0\,P_N(n,t|n_0)\,P_N^\star(n_0)-\mu_N^2
  }{\sigma_N^2}
  \nonumber\\
    &=&
%  \textcolor{red}{
    \textrm{(using Eq.~\eqref{eq_morse_identity})}
%  }
  \nonumber\\
  &=&
%  \textcolor{red}{
    \dfrac{1}{\sigma_N^2}
    \int_0^{2\pi}\mathrm{d}\theta
    \dfrac{
      \mathrm{e}^{-[(1+g)-2\sqrt{g}\cos(\theta)]\,\mu t}
    }{
      \left[(1+g)-2\sqrt{g}\cos(\theta)\right]
    }\,
    \sum_{n=0}^\infty\sum_{n_0=0}^\infty
    n\,n_0\,(1-g)\,g^{n_0}\,\dfrac{g^{\frac{n-n_0}{2}}}{\pi}\cdot
%  }
  \nonumber\\
  &&\qquad
 % \textcolor{red}{
    \cdot\left[\sin(n_0\theta)-\sqrt{g}\sin((n_0+1)\theta)\right]
    \,\left[\sin(n\theta)-\sqrt{g}\sin((n+1)\theta)\right]
%  }
    \nonumber\\
  &=&
%    \textcolor{red}{
      \dfrac{(1-g)^2}{g}
      \,\int_0^{2\pi}\mathrm{d}\theta
      \dfrac{
        \mathrm{e}^{-[(1+g)-2\sqrt{g}\cos(\theta)]\,\mu t}
      }{
        \left[(1+g)-2\sqrt{g}\cos(\theta)\right]
      }\,
      \dfrac{
        g\,(1-g)\,\sin^2(\theta)
      }{
        \pi\,\left[(1+g)-2\sqrt{g}\cos(\theta)\right]^2
      }
%    }
  \nonumber\\
  &=&
  \dfrac{(1-g)^3}{\pi}\int_0^{2\pi}\mathrm{d}\theta
  \dfrac{
    \sin^2(\theta)
    \,\mathrm{e}^{-[(1+g)-2\sqrt{g}\cos(\theta)]\,\mu t}
  }{
    \left[(1+g)-2\sqrt{g}\cos(\theta)\right]^3
  }
  \nonumber\\
  &\simeq&
  \mathrm{e}^{-t/\tau}\,,
\end{eqnarray}
with
\begin{equation}
  \tau=\dfrac{1+g}{(1-g)^2\,\mu}\,.
\end{equation}
The latter approximation is obtained imposing
$\displaystyle\int_0^\infty\mathrm{d}t\,\rho(t)=\int_0^\infty\mathrm{d}t\,\mathrm{e}^{-t/\tau}$.

%\subsubsection{Short-time expansion}
\section{Short-time expansion}
From Eq.~\eqref{eq_morse_solution}  and the identity
\mbox{$\int_0^{2\pi}\mathrm{d}\theta\,\sin(n_0\,\theta)\,\sin(n\,\theta)=\pi\,\delta_{n,n_0}\,\mathrm{U}(n_0-1)$}
we also get a useful short time
expansion: 
\begin{equation}
  P_N(n,t|n_0)
  =\left\{
    \mathrm{U}(n)\,-\mu t
    \,[\mathrm{U}(n-1)+\mathrm{U}(n)\,g]
  \right\}
  \delta_{n,n_0}
  +\mu t\,
  \left[\mathrm{U}(n)\,\delta_{n+1,n_0}+g\,\mathrm{U}(n-1)\,\delta_{n-1,n_0}
    \right]
  +\mathcal{O}(t^2)\,.
    \label{eq_short_time_expansion}
\end{equation}
Inserting this expansion in
\begin{equation}
  \mathbb{E}[(\boldsymbol{R}_{\mathrm{CM}}(t)-\boldsymbol{R}_{\mathrm{CM}}(0))^2|n_0]
  =3\,\mathbb{E}[S(t)|n_0]
  =3\sum_{n=0}^\infty\dfrac{2\,D_0}{(n+3)^\alpha}\,
  \int_0^t\mathrm{d}t'P_N(n,t'|n_0)\,,
\end{equation}
we obtain the following
leading terms for short time evolution of the squared displacement of
the polymer CM
\begin{eqnarray}
  \label{eq_brownian_1}
  \mathbb{E}[(\boldsymbol{R}_{\mathrm{CM}}(t)-\boldsymbol{R}_{\mathrm{CM}}(0))^2|n_0]
  &=&6\,\dfrac{D_0}{(n_0+3)^\alpha}
  \left[
    \left(t-\dfrac{\mu\,t^2}{2}g\right)\mathrm{U}(n_0)-\dfrac{\mu\,t^2}{2}\mathrm{U}(n_0-1)
    \right]
  \nonumber\\
  &&
  +\,6\,\dfrac{D_0}{(n_0+2)^\alpha}
  \dfrac{\mu\,t^2}{2}\,\mathrm{U}(n_0-1)
  \nonumber\\
  &&
  +6\,\dfrac{D_0}{(n_0+4)^\alpha}\,\dfrac{\mu\,t^2}{2}\,g\,\mathrm{U}(n_0)
  +\mathcal{O}(t^3)\,.
  \nonumber
\end{eqnarray}

%\subsection{Gillespie simulations}
\section{Langevin-Gillespie simulations}
\label{sec_gillespie}
We compare Gillespie
simulations~\cite{gillespie1977} of the polymerization process $N(t)$ and the
parallel Langevin dynamics of the polymer CM,
$\boldsymbol{R}_{\mathrm{CM}}(t)=(X_{\mathrm{CM}}(t),Y_{\mathrm{CM}}(t),Z_{\mathrm{CM}}(t))$,
with analytical results.

As reported in the main text, from
polymer physics~\cite{deGennes1979,Doi1992}
it is known that the CM of a polymer chain with
$N+n_{\mathrm{min}}$ subunits diffuses with
$D(N)=D_{0}/(N+n_{\mathrm{min}})^{\alpha}$, $D_{0}$ being a diffusion coefficient
characteristic of the filament and $\alpha$
specific to the environment conditions ($\alpha=1/2, 1,2$ for Zimm,
Rouse,  and reptation, respectively).
Given the number of monomers $N(t)+3=n(t)+3$ in the polymer
chain, e.g. the coordinate $x_{\mathrm{CM}}$
updates thus as  
\begin{equation}
  x_{\mathrm{CM}}(t+\mathrm{d}t) = x_{\mathrm{CM}}(t)
  + \sqrt{\dfrac{D_0}{(n(t)+3)^\alpha}\,\mathrm{d}t}\;\;\mathcal{N}(0, 1),
\end{equation}
where $\mathcal{N}(0,1)$ is the Gaussian distribution with zero mean and unit variance.

Concomitantly, the stochastic variable $N(t)$ undergoes the $M/M/1$
(Markovian interarrivaltimes/Markovian service times/1
server)~\cite{Jain2007} birth-death process, which is efficiently
simulated through the Gillespie algorithm~\cite{gillespie1977}.
Given $N(t)=n(t)$, the $n$ update equation complies with: 
\begin{itemize}
\item if $n(t)>0$, then
  \begin{itemize}
  \item[*] $n(t+\mathrm{d}t)=n(t)+1$ with probability
    $\lambda\,\mathrm{d}t=g\,\mu\,\mathrm{d}t$,
  \item[*] $n(t+\mathrm{d}t)=n(t)-1$ with probability
    $\mu\,\mathrm{d}t$,
  \item[*] $n(t+\mathrm{d}t)=n(t)$ with probability $1-(1+g)\,\mu\,\mathrm{d}t$;
  \end{itemize}
\item if $n(t)=0$, then
  \begin{itemize}
  \item[*] $n(t+\mathrm{d}t)=n(t)+1$ with probability
    $\lambda\,\mathrm{d}t=g\,\mu\,\mathrm{d}t$,
  \item[*] $n(t+\mathrm{d}t)=n(t)$ with probability $1-g\,\mu\,\mathrm{d}t$.
  \end{itemize}
\end{itemize}

Once an ensemble of simulations has been generated,
probability density functions (PDFs)
and moments of the stochastic
process $\boldsymbol{R}_{\mathrm{CM}}(t)$ can then be numerically inferred.
The following plots summarize comparisons between numerical and
analytical results with
$\mathrm{d}t/\tau=0.1$.
\begin{figure}[t]
  \begin{center}
    \includegraphics[width=0.5\columnwidth]{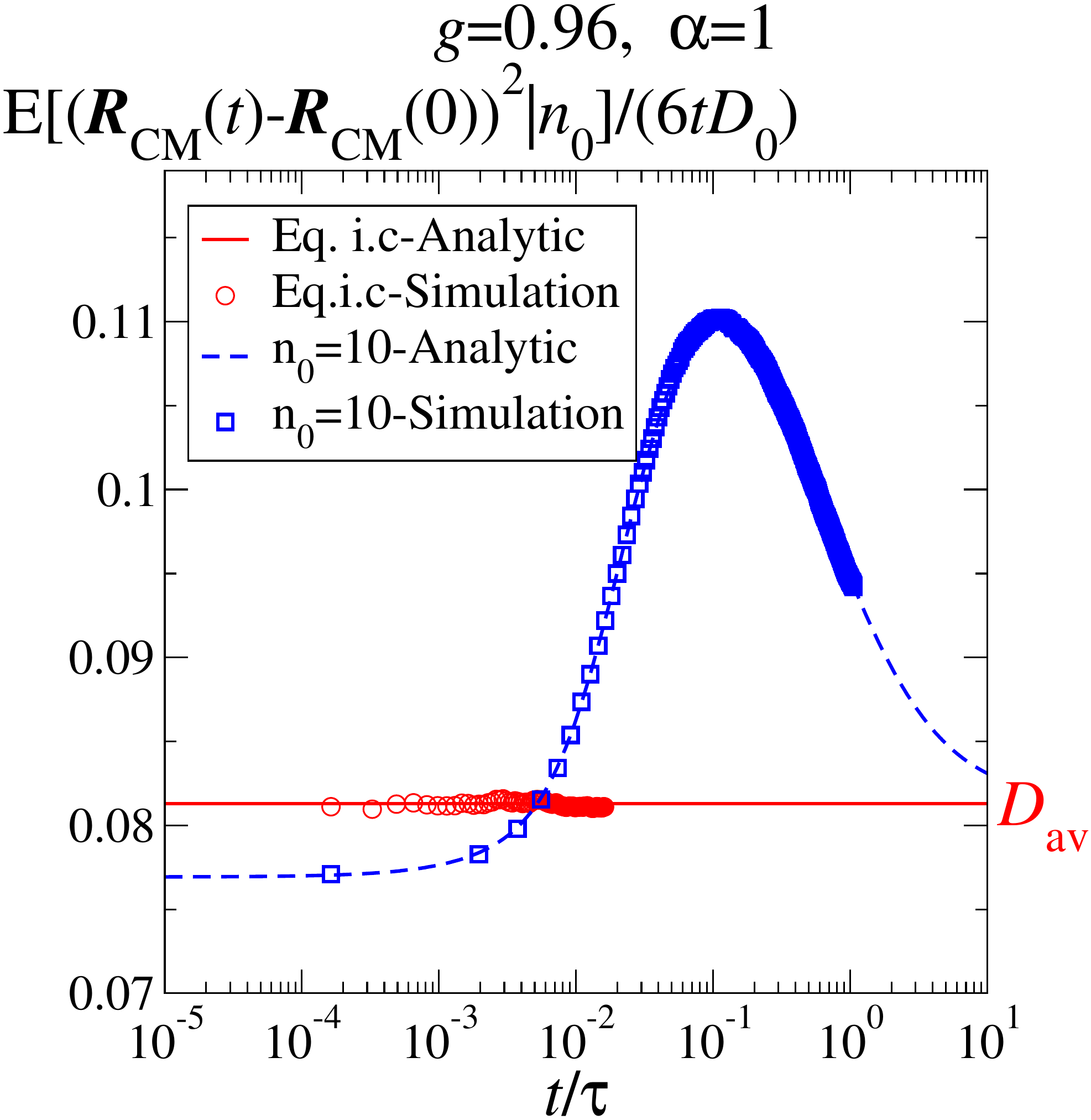}\\
  \end{center}
  \caption{Time evolution of the CM second moment obtained through
    Gillespie-Langevin simulations (symbols) contrasted with analytical results for both
    equilibrium and nonequilibrium initial polymer sizes. Simulation
    data are obtained averaging over $7\times10^{5}$ ($n_0=10$) and
    $8\times10^{5}$ realizations (equilibrium).}
\end{figure}

\begin{figure}[t]
  \begin{center}
    \includegraphics[width=0.5\columnwidth]{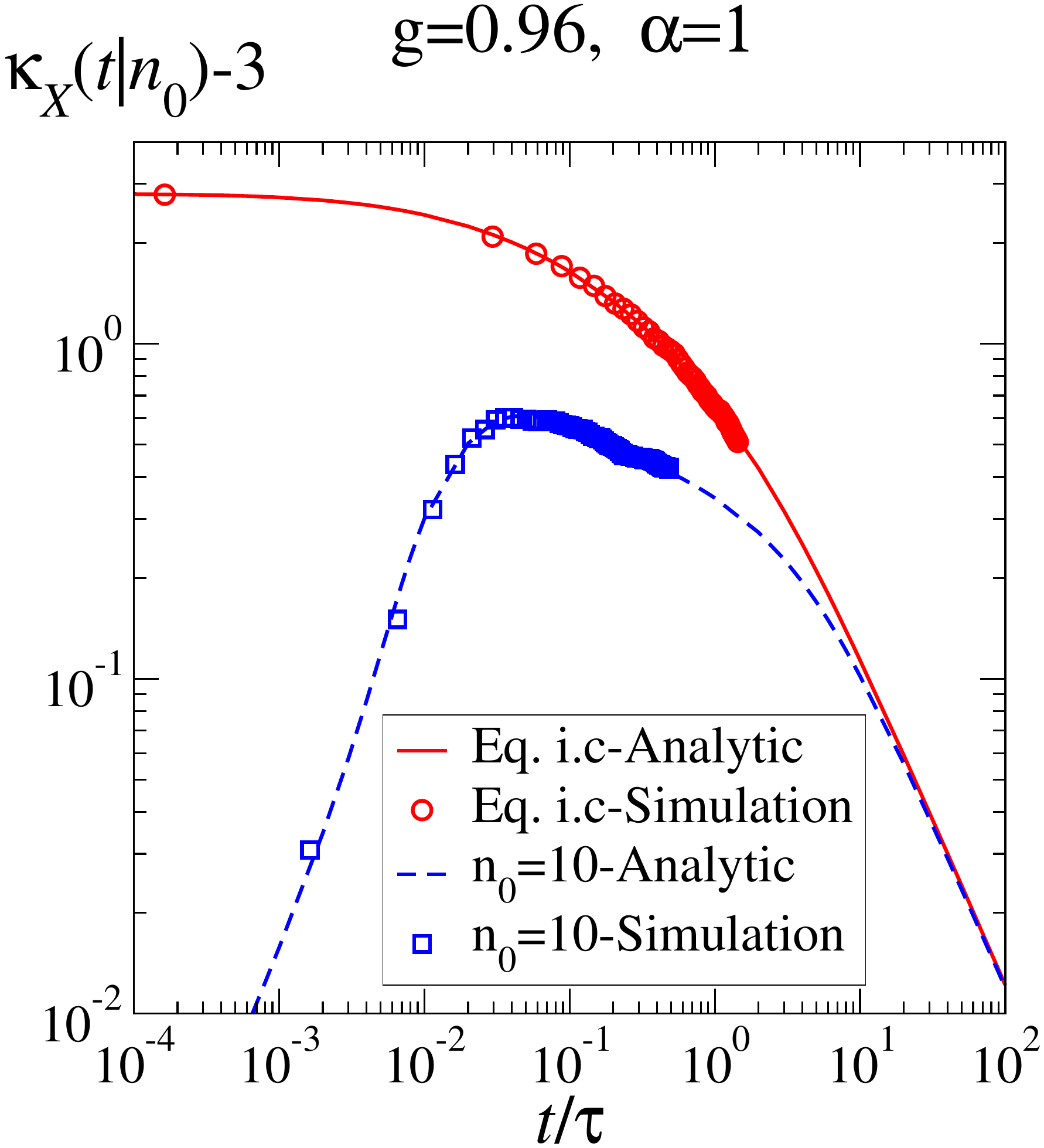}\\
  \end{center}
  \caption{Time evolution of the CM $x$-kurtosis obtained through
    Gillespie-Langevin simulations (symbols) contrasted with analytical results for both
    equilibrium and nonequilibrium initial polymer sizes. Simulation
    %\textcolor{red}{
      data
    %}
    are obtained averaging over $9\times10^{5}$ ($n_0=10$) and
    $4\times10^{5}$ realizations (equilibrium).}
\end{figure}

\begin{figure}[t]
  \begin{center}
    \includegraphics[width=0.5\columnwidth]{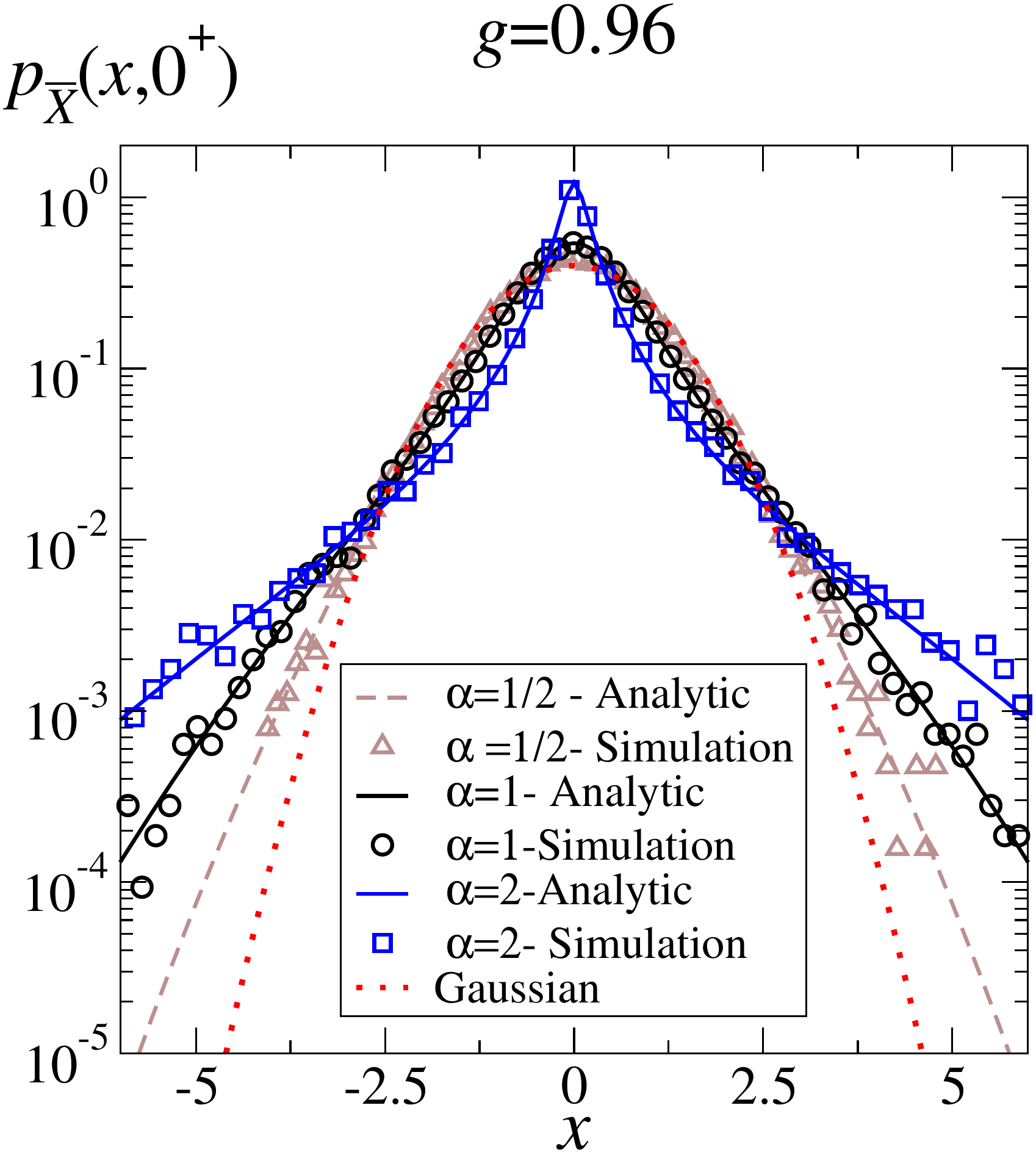}\\
  \end{center}
  \caption{
    Unit-variance initial $x$-PDF for the CM of
    %\textcolor{red}{
      different polymer models
      %}
    with equilibrium sizes $P_N(n_0,0)=P^\star_N(n_0)$.
    Gillespie-Langevin simulations (symbols) are contrasted with
    analytical results. Simulation
    %\textcolor{red}{
      data
    %}
    are obtained averaging over $4\times10^{5}$ realizations.
    For comparison purposes, a unit-variance Gaussian
    PDF is also plotted in red dotted line.
  }
  \label{fig_initial_pdf_sm}
\end{figure}

\section*{References}
%\begin{thebibliography}{<num>}
\bibliography{draft_bibliography}
%\end{thebibliography}

\end{document}